\documentclass[twocolumn]{revtex4}

\input epsf
\usepackage{amsmath,amssymb}
\usepackage{graphicx}

\draft

\begin{document}
\titlepage
\title{Can vacuum decay in our Universe?}
\author{Peng Wang$^1$\footnote{E-mail: pewang@eyou.com} and Xin-He Meng$^{1,2,3}$ \footnote{E-mail: xhm@physics.arizona.edu}}

\affiliation{1. Department of Physics, Nankai University, Tianjin,
300071, P.R.China \\ 2. Department of Physics, University of
Arizona, Tucson, AZ 85721, USA \\3. CCAST (World Lab.), P.O.Box
8730, Beijing 100080, P.R.China}

\begin{abstract}
We take a phenomenological approach to study the cosmological
evolution of decaying vacuum cosmology ($\Lambda(t)$CDM) based on
a simple assumption about the form of the modified matter
expansion rate. In this framework, almost all the current vacuum
decaying models can be unified in a simple manner. We argue that
the idea of letting vacuum decay to resolve the fine-tuning
problem is inconsistent with cosmological observations. We also
discuss some issues in confronting $\Lambda(t)$CDM with
observation. Using the effective equation of state formalism, we
indicate that $\Lambda(t)$CDM is a possible candidate for phantom
cosmology. Moreover, confronted with a possible trouble of
effective equation of state formalism, we construct the effective
dark energy density. Finally, we discuss the evolution of linear
perturbation.
\end{abstract}

\maketitle

\section{Introduction}
\label{intro}

The cosmological constant problem has a long and checkered history
(see Refs.\cite{carroll-cc, weinberg} for reviews). The ``old"
cosmological constant problem is to understand why the vacuum
energy is so small. In light of the recent type Ia supernova
observation \cite{obs}, the new cosmological constant problem is
to understand why the vacuum energy is not only small, but also of
the same order of magnitude as the present matter density of the
Universe. This is much harder than the old one (so a lot of
alternative explanations of the cosmic acceleration have appeared
in the literature. See Refs.\cite{carroll-de, peebles} for
reviews).

Allowing the cosmological constant to be dynamical to resolve the
fine-tuning has been proposed rather early \cite{bron} and the
cosmological consequence of various specific models have been
discussed extensively in recent years \cite{carvalho, H, vish,
arbab, overduin, ray, taha, bertolami} and references therein. It
is worth mentioning that some authors have also argued that a
dynamical cosmological constant is the requirement of the laws of
Quantum Field Theory \cite{Panangaden, odintsov, wagoner, cohen,
shapiro, renorm, wein} (see Ref.\cite{milton, Padmanabhan} for
recent reviews).

To get a definite model of $\Lambda(t)$CDM cosmology, one should
specify a vacuum decay law. There are a lot of proposals of the
vacuum decay law in the literature (see Ref.\cite{overduin} for a
comprehensive list). We will review shortly five of them whose
cosmological implications are mostly discussed in the literature.

I. $\Lambda\propto H^2$, where $H^2=\dot a/a$ is the Hubble
parameter. This was proposed by Carvalho et al. \cite{carvalho}
based on dimensional argument (see Ref.\cite{H} for subsequent
discussions). The same law can also be deduced by argument related
to effective field theory and black hole thermodynamics
\cite{cohen, hsu}. Recently, it is argued in Ref.\cite{hsu} that
this type of decaying law is inconsistent with current constraint
on the dark energy equation of state. While we agree with this
conclusion, we do not agree with the author's argument. We will
present our argument disfavoring this decaying law in
Sec.\ref{evolution}. See also Ref.\cite{Horvat} for related
discussion.

II. $\Lambda\propto R=6(H^2+\ddot a/a)$ where $R$ is the scalar
curvature. This was proposed by Al-Rawaf and Taha \cite{taha} in
order to solve the entropy problem.

III. $\Lambda\propto (\ddot a/a)$. This was proposed by Arbab
\cite{arbab}, which has used an argument based on the law
$\Lambda\propto R$.

IV. $\Lambda\propto \rho_m$, where $\rho_m$ is the matter energy
density. This was proposed by Vishwakarma \cite{vish} using
dimensional argument.

V. $d\Lambda/dz\propto dH^2/dz$. This was proposed by Shapiro and
Sola \cite{shapiro} based on renormalization group argument and
its cosmological constraint is recently discussed in detail in
Ref.\cite{renorm}.

Recently, Ray and Mukhopadhyay \cite{ray} showed that the laws I,
III and IV are actually dynamically equivalent. More concretely,
if we denote the dimensionless proportional constants in those
three models as $3\alpha$, $\beta$, $\gamma8\pi G$, then the three
laws will give precisely the same cosmological evolution under the
condition that
\begin{equation}
\alpha={\beta\over 3(\beta-2)}={\gamma\over 1+\gamma}\ .\label{}
\end{equation}
Note that we have assumed that besides vacuum energy, the
remaining component of the Universe is mainly pressureless matter.
We think this equivalence actually means that we cannot
distinguish those three laws by cosmological observations! So in
view of cosmology, there is actually no need to discuss them
separately and if one of them is excluded by observation, all the
remaining one are also excluded.

This observation motivates us to explore in a more general ground
the equivalence between different vacuum decaying laws.
Complementarily, we also want to ask the question: whether there
are some general features in decaying vacuum cosmology that are
independent of the decaying law? If this is true, we can
faithfully talk about constraining the vacuum decaying rate even
if we still do not understand the physics underlying the law.

In the usual way of studying $\Lambda(t)$CDM, one need first
specify a vacuum decay law, then the evolution of CDM and the
vacuum energy can be solved. In this paper, we will take a
different approach. The starting point of this work is an
assumption about the form of the modification of the CDM expansion
rate due to the vacuum decay (see Eq.(\ref{6}) below). We think
this is a rather physical approach since in cosmological
observations, the vacuum decay rate actually cannot be directly
observed and it is the modified CDM expansion rate that we can
directly observe (if it indeed exists). In other words, we can
prove or disprove a particular vacuum decay law by cosmological
observation only through its effects on the CDM expansion rate. We
will try to probe what features of the vacuum decay law can be
deduced from this simple assumption and how can we constrain them.
In this approach, we can also probe the question of whether we can
discriminate between various vacuum decay laws by cosmological
observations. Roughly speaking, what we want to do can be said as
a ``model-independent" study of decaying vacuum cosmology.

\section{Cosmological evolution of $\Lambda(t)$CDM}
\label{evolution}

We will be mainly interested in the late cosmological evolution,
so we will consider only the case of the vacuum energy decaying to
cold dark matter (CDM). From the equation $T^{\mu\nu}_{\ \
;\nu}=0$, we can find that the standard continuity equation for
CDM will be modified by a term that is proportional to the vacuum
energy decay rate,
\begin{equation}
\dot\rho_m+3H\rho_m=-\dot\rho_{\Lambda}\ ,\label{5}
\end{equation}
where $\rho_{\Lambda}\equiv\Lambda/8\pi G$ is the energy density
of vacuum. Note that although the vacuum is decaying, the physical
equation of state (EOS) of the vacuum $\omega_\Lambda\equiv
p_\Lambda/\rho_\Lambda$ is still equal to constant $-1$, which
follows from the definition of the cosmological constant.

Since vacuum energy is constantly decaying into CDM, CDM will
dilute in a smaller rate compared with the standard relation
$\rho_m\propto a^{-3}$. Thus we assume that the energy density of
the CDM will dilute in a rate whose deviation from the standard
case can be characterized by a positive small \emph{constant}
$\epsilon$,
\begin{equation}
\rho_m=\rho_{m0}a^{-3+\epsilon}\ . \label{6}
\end{equation}
where $\rho_{m0}$ is the present value of $\rho_m$. This is the
main assumption of this paper, which we think is rather natural.
Actually, this assumption is valid in all the existing models of
$\Lambda(t)$CDM. An immediate simple observation is that we must
have $\epsilon \le1$. Otherwise the Universe will expand
accelerated in the matter dominated era, which is excluded by the
observation of SNe Ia that our Universe expanded decelerated
before redshift $z\sim 0.5$ \cite{riess}. Actually, we should
expect $\epsilon\ll 1$ since so far there has been no report from
observation about an anomalous CDM expansion rate.

Rewriting the continuity equation (\ref{5}) in the form
\begin{equation}
{d\rho_m\over da}+3{\rho_m\over a}=-{d\rho_{\Lambda}\over da}\
,\label{con2}
\end{equation}
then substituting the ansatz (\ref{6}) into Eq.(\ref{con2}), we
can find that $\rho_{\Lambda}$ is given by
\begin{equation}
\rho_{\Lambda}=\tilde\rho_{\Lambda0}+{\epsilon\rho_{m0}\over
3-\epsilon}a^{-3+\epsilon}\ ,\label{6.1}
\end{equation}
where $\tilde\rho_{\Lambda0}$ is an integration constant
representing the ground state value of the vacuum. Note that the
present value of the vacuum energy density is
$\rho_{\Lambda0}=\tilde\rho_{\Lambda0}+{\epsilon\over
3-\epsilon}\rho_{m0}$.

It is interesting to observe that the decaying part of the vacuum
energy, i.e. the second term in Eq.(\ref{6.1}), dilutes at exactly
the same rate as that of CDM. Actually, it is easy to see from
Eq.(\ref{con2}) that the vacuum always dilutes in the same rate as
the dominate fluid of the Universe if the fluid dilutes in a power
law form. So the vacuum decay rate always exhibits a ``tracking
behavior". As will be discussed in more detail below, it is just
this tracking behavior that makes a $\Lambda(t)$CDM model with
zero vacuum ground state unrealistic model of our Universe.

Based on the prediction of inflation \cite{liddle} and the recent
data from WMAP \cite{wmap}, we will assume spatial flatness in
this paper. Thus the Friedamnn equation of $\Lambda(t)$CDM reads,
\begin{eqnarray}
H^2&=&{8\pi G\over3}(\rho_m+\rho_{\Lambda})\cr
&=&H_0^2\left({3\Omega_{m0}\over
3-\epsilon}(1+z)^{3-\epsilon}+\tilde\Omega_{\Lambda0}\right)\
,\label{7}
\end{eqnarray}
where $\Omega_{m0}\equiv \rho_{m0}/\rho_{c0}$,
$\tilde\Omega_{\Lambda0}\equiv \tilde\rho_{\Lambda0}/\rho_{c0}$
and $\rho_{c0}=3H_0^2/8\pi G$ is the present critical energy
density. Note that by the assumption of spatial flatness, we have
$\tilde\Omega_{\Lambda0}=1-3\Omega_{m0}/(3-\epsilon)$. From the
Friedmann equation (\ref{7}), the Universe will expand
exponentially when the term $\tilde\rho_{\Lambda0}$ dominates.
When the first term dominates, i.e. matter domination, the scale
factor evolves like, $a\propto t^{2/(3-\epsilon)}$ and $H=2/
(3-\epsilon)t$. Thus the Universe will expand slower during matter
domination than standard $\Lambda$CDM and thus the Universe
becomes older in $\Lambda(t)$CDM.

The deceleration equation follows from Eqs.(\ref{7}) and
(\ref{5}),
\begin{eqnarray}
{\ddot a\over a}&=&-{4\pi G\over3}(\rho_m-2\rho_\Lambda)\cr
&=&-{4\pi
G\over3}\left({3-3\epsilon\over3-\epsilon}\rho_{m0}(1+z)^{3-\epsilon}-2\tilde\rho_{\Lambda0}\right)
\ .\label{dece}
\end{eqnarray}

\begin{figure}
  \includegraphics
  [width=0.8\columnwidth]{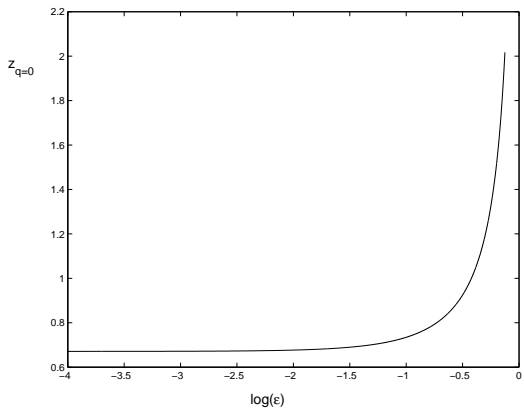}
  \caption{The dependence of the turning redshift $z_{q=0}$ on $\epsilon$ given by Eq.(\ref{turnpt}) for $\Omega_{m0}=0.3$.}\label{1}
\end{figure}

From this we can find the parameter $z_{q=0}$ which is defined to
be the zero point of the deceleration parameter $q(z_{q=0})=0$.
From Eq.(\ref{dece}), it is given by,
\begin{equation}
z_{q=0}=\left [{6-2\epsilon\over3-3\epsilon}\left
(\Omega_{m0}^{-1}-{3\over3-\epsilon}\right)\right
]^{{1\over3-\epsilon}}-1\ .\label{turnpt}
\end{equation}

Fig.1 shows the dependence of $z_{q=0}$ on $\epsilon$. From it we
can see that $z_{q=0}$ will be larger when $\epsilon$ is larger,
i.e. the Universe begins to accelerate earlier for larger
$\epsilon$. We can also see that when $\epsilon<0.1$, $z_{q=0}$
converge to the value in the standard $\Lambda$CDM case.

Equations (\ref{6}), (\ref{6.1}), (\ref{7}) and (\ref{dece})
determine completely the evolution of a Universe in which the
vacuum is decaying. It is only the result of the assumption
(\ref{5}) and the laws of General Relativity. So we can see that
under the assumption (\ref{5}), the most general $\Lambda(t)$CDM
cosmology can be characterized by only two parameters:
$\tilde\rho_{\Lambda0}$ and $\epsilon$. Since the assumption
(\ref{5}) is also valid in all the existing models of
$\Lambda(t)$CDM, we should expect that all the existing models of
$\Lambda(t)$CDM can be reproduced by special choice of the
parameters $\tilde\rho_{\Lambda0}$ and $\epsilon$. This is indeed
the case. Model I with proportional constant $3\alpha$ is just
given by $\tilde\rho_{\Lambda0}=0$ and $\epsilon=3\alpha$. Model
III with proportional constant $\beta$ is just given by
$\tilde\rho_{\Lambda0}=0$ and $\epsilon=\beta/(\beta-2)$. Model IV
with proportional constant $\gamma8\pi G$ is just given by
$\tilde\rho_{\Lambda0}=0$ and $\epsilon=3\gamma/(1+\gamma)$. Thus
in our framework, we easily reproduced the result of
Ref.\cite{ray} that those three laws are dynamically equivalent.
Model II with proportional constant $\zeta$ is just given by
$\tilde\rho_{\Lambda0}=0$ and $\epsilon=3\zeta/(1-\zeta)$. Thus we
can see that actually the all the models I, II, III and IV are
dynamically equivalent. Specifically, in those four models, the
ground state of the vacuum must be zero. Model V with proportional
constant $3\nu$ is just given by $\epsilon=3\nu$ and
$\tilde\rho_{\Lambda0}$ an arbitrary value. The freedom in
choosing $\tilde\rho_{\Lambda0}$ is conceivable since in this law
$\Lambda$ appears as a derivative and so is constrained up to a
constant. Thus we can see that cosmology in renormalization group
model is just the most general cosmology for a vacuum decaying
Universe.

It can be seen from Eqs.(\ref{6.1}) and (\ref{7}) that if the
ground state of the vacuum is zero, i.e.
$\tilde\rho_{\Lambda0}=0$, as in models I, II, III and IV, we can
get a currently accelerating Universe only if $\epsilon>1$.
However, this is unacceptable because this means the Universe will
also be accelerating in the matter dominated era from
Eq.(\ref{6}). Furthermore, in this case, the ratio of vacuum
energy density $\rho_\Lambda$ and the CDM energy density $\rho_m$
will be a constant $\rho_m/\rho_\Lambda=(3-\epsilon)/\epsilon$.
This means that the Universe can never change from a matter
dominated era to a vacuum dominated era. So the Universe is either
always accelerating or always decelerating from the onset of
matter domination to today. Both cases are excluded by
observation. Thus we conclude that those four models is
unrealistic models of our Universe. This conclusion agrees with
that of Ref.\cite{hsu}. In Ref.\cite{hsu}, the author also noticed
the tracking behavior of the vacuum energy, then from this he
concluded that vacuum will dilute as $a^{-3}$ and then due to the
standard relation $\rho\propto a^{-3(1+\omega)}$, this will give a
vacuum EOS equaling to zero that is incompatible with current
constraint on dark energy EOS. We do not agree with this argument.
First, from Eq.(\ref{6.1}), vacuum will not dilute as $a^{-3}$ due
to the interaction between vacuum and matter. Second, in
$\Lambda(t)$CDM, the standard relation $\rho\propto
a^{-3(1+\omega)}$ between energy density and EOS fails for both
matter and vacuum energy which is due to the modified continuity
equation (\ref{5}). Third, by choosing $\epsilon=3$, the Universe
will expand exponentially and thus model I can fit perfectly with
current constraint on dark energy. The real reason that model I is
incompatible with observation is that this model cannot
accommodate simultaneously an accelerating Universe today and a
decelerating Universe during matter dominated era.

So in order for $\Lambda(t)$CDM to give realistic cosmology, we
must have $\tilde\rho_{\Lambda0}$ to be non-zero and to be of the
order $(10^{-3} eV)^4$. We think this means that \emph{in
$\Lambda(t)$CDM we cannot evade the fine-tuning problem in
ordinary $\Lambda$CDM cosmology.} The question of ``why the vacuum
energy is so small" changes to ``why the ground state value of the
vacuum energy is so small", which we think is not easier or
technically more natural than the original one.

It is worth mentioning that in order to circumvent the above
difficulty. A possible way is assuming that the rate of vacuum
decaying into baryon and CDM is different. Imagine that
$\epsilon_{baryon}$ is smaller than one and $\epsilon_{CDM}$ is
larger than one. Then, if CDM dominates over baryon only recently,
we can have a decelerating Universe in the past and an
accelerating Universe in the present. However, we think this
cannot work. Since this requires baryon dominates over the CDM
when redshift is larger than one so the Universe is actually
baryon-dominated after recombination to the formation of large
scale structure. As is well-known, this cannot reproduce the
observed large scale structure. Actually, the problem of large
structure formation in baryon-dominated Universe is worsen in
decaying vacuum cosmology because evolution of linear perturbation
will be slowed down if vacuum is decaying, see Eq.(\ref{per}).

Finally, from Eqs.(\ref{6.1}), (\ref{7}) and (\ref{dece}), we can
find that the following vacuum decaying law can also give a
$\Lambda(t)$CDM model with arbitary $\tilde\rho_{\Lambda0}>0$
\begin{equation}
{d\Lambda\over da}={\epsilon\over 24\pi G(1+\epsilon)}{dR\over
da}\ ,\label{law}
\end{equation}
where $R$ is the scalar curvature of the spacetime. Up to our
knowledge, this form of vacuum decaying law has not been discussed
in the literature. We think one of its appealing feature is that
all the quantities appearing in it is covariant. Now we still do
not know whether this can really be predicted from a fundamental
theory. But this is surely an interesting question that deserves
further investigation. Note that based on our above discussion,
the law (\ref{law}) cannot be distinguished from the model V by
cosmological observations. Their difference lies only in that
whether they can be derived from a fundamental theory.

\section{Confronting $\Lambda(t)$CDM with observation} \label{effectiveeos}

In this section we discuss the observational consequence of the
system (\ref{5}-\ref{dece}). Since they are essentially the same
as the one in renormalization group model \cite{renorm}, some of
its conclusions can be applied here. Specifically, they performed
data fitting with SNe Ia observation and considered constraint
from BBN, they found the constraint $\epsilon<0.3$. So in our
discussion, we will focus our attention to the case of
$\epsilon<0.1$ in numerical illustrations.

However, due to the modified expansion rate, confronting
$\Lambda(t)$CDM with observation is a subtle issue: care should be
paid to understand what are the real parameters in $\Lambda(t)$CDM
that is constrained in various cosmological observations. For
example, we \emph{cannot} compare the vacuum energy density
(\ref{6.1}) or the physical EOS $\omega_\Lambda=-1$ directly to
current constraints on the dark energy density \cite{ywang} and
EOS \cite{ywang, omega}. Recall that in constraining dark energy
EOS, we take the standard relation $\rho_m\propto a^{-3}$ for CDM
and assume \emph{a prior} an ansatz for the dark energy EOS
\cite{omega} or energy density \cite{ywang}, then we constrain the
parameter(s) in the ansatz using data from SNe Ia, CMB and LSS
observation. So the constraint on dark energy EOS in the
literature \cite{ywang, omega} depends crucially on the assumption
of standard expansion rate of the CDM.

Due to the modified matter expansion rate, it is the effective EOS
defined in Ref.\cite{linder} that is the real quantity in
$\Lambda(t)$CDM that is constrained in dark energy reconstruction
works. Rewrite the Friedmann equation (\ref{7}) in the form
\begin{equation}
{H^2(z)\over H_0^2}=\Omega_{m0}(1+z)^3+{\delta H^2(z)\over H_0^2}\
,\label{1}
\end{equation}
where $\delta H$ is given by
\begin{equation}
{\delta H^2(z)\over H_0^2}={3\Omega_{m0}\over
3-\epsilon}(1+z)^{3-\epsilon}-\Omega_{m0}(1+z)^3+\tilde\Omega_{\Lambda0}\
.\label{7.1}
\end{equation}
Note that the term $\delta H^2$ should be interpreted as
$\delta(H^2)$, not $(\delta H)^2$, so it can take both positive
and negative value. For example, in $\Lambda(t)$CDM, as can be
seen from Eq.(\ref{7.1}), $\delta H^2$ is positive at small
redshift and negative at large redshift.

The effective EOS $\omega_{X}$ is defined as \cite{linder}
\begin{equation}
\omega_{X}=-1+{1\over 3}\frac{d\ln\delta H^2(z)}{d\ln(1+z)}\
,\label{3}
\end{equation}
while the explicit form in $\Lambda(t)$CDM can be found from
Eq.(\ref{7.1}),
\begin{equation}
\omega_{X}=-1+\frac{(1+z)^{3-\epsilon}-(1+z)^3}{{3\over
3-\epsilon}(1+z)^{3-\epsilon}-(1+z)^3+\tilde\Omega_{\Lambda0}/\Omega_{m0}}\label{eos}
\end{equation}

It is worth mentioning that this definition of effective EOS is
motivated by the fact that from the definition (\ref{3}), the
Friedmann equation (\ref{7}) can be written in the form
\begin{eqnarray}
{H^2(z)\over H_0^2}&=&\Omega_{m0}(1+z)^3+(1-\Omega_{m0})\times\cr
&\times&\exp\left(3\int^z_0d\ln(1+z')[1+\omega_X(z')]\right)\
.\label{0}
\end{eqnarray}
This is just the standard form of the Friedmann equation in which
case the dark energy is modelled as an ideal fluid with equation
of state $\omega_X$. Thus $\omega_X$ is the real quantity in
$\Lambda(t)$CDM that is constrained in dark energy reconstruction
works.

\begin{figure}
  \includegraphics[width=0.8\columnwidth]{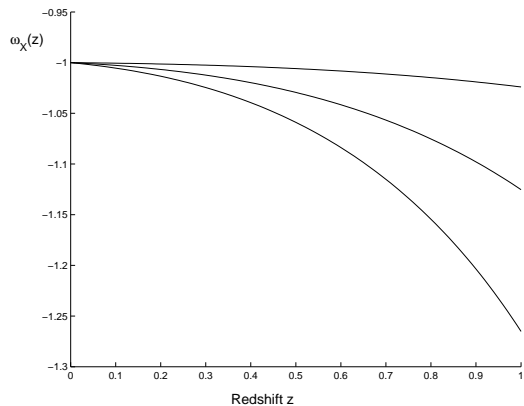}
  \caption{The effective EOS given by Eq.(\ref{eos}). The three lines correspond to $\epsilon=
  0.01, 0.05, 0.1$ from top to bottom.}\label{1}
\end{figure}

Fig.2 shows the evolution of $\omega_X$ up to redshift 1 for
$\epsilon=0.01,  0.05, 0.1$ from top to bottom. It is easy to see
that $\omega_{X}<-1$ for small $z$ while $\omega_X(z=0)=-1$ and if
$\epsilon\ll1$, $\omega_X$ changes slowly with time. In current
works of constraining $\omega_{X}$, the most reliable results
assumed that $\omega_{X}$ is a constant \cite{omega, ywang} (this
is conceivable since allowing EOS to vary will increase the number
of parameters to constrain). Thus what is actually constrained in
those works is the average value of the effective EOS
$\bar\omega_{X}\equiv\int\Omega_X(a)\omega_X(a)da/\int\Omega_X(a)da$
\cite{limin}, which from Eq.(\ref{eos}) is smaller than $-1$ if
averaged over small redshift. Thus $\Lambda(t)$CDM is a possible
candidate of ``phantom energy" \cite{caldwell}. In this picture,
$\bar\omega_X<-1$ is not the result of exotic behavior of the dark
energy, but the modified expansion rate of the CDM, i.e. decaying
vacuum can trick us into thinking that $\omega_X<-1$
\cite{carroll-trick} (See Ref.\cite{Onemli} for another
interesting model of this type). This possibility would be more
interesting when considering the fact that modelling dark energy
as scalar field with $\omega_X<-1$ has encountered fundamental
theoretical difficulties \cite{carroll-phantom}. Furthermore, in
$\Lambda(t)$CDM, the argument leading to ``Big Rip" fate of the
Universe in phantom cosmology \cite{kamion} no longer applies. The
Universe will just expand exponentially forever just as in
$\Lambda$CDM. Under the assumption of constant EOS, current
constraint on $\bar\omega_X$ from SNe Ia data is
$-4.36<\bar\omega_X<0.8$ at $95\%$ C.L. \cite{ywang}. Then
$\epsilon<0.1$ is obvious consistent with current constraint on
dark energy EOS. Moreover, it is easy to see from Fig.2 that
$\bar\omega_X$ will be more and more negative when the redshift
gets larger. So if more high redshift SNe Ia data is available in
the future, $\Lambda(t)$CDM predicts that we will get a more
negative $\bar\omega_X$.

\begin{figure}
  \includegraphics[width=0.8\columnwidth]{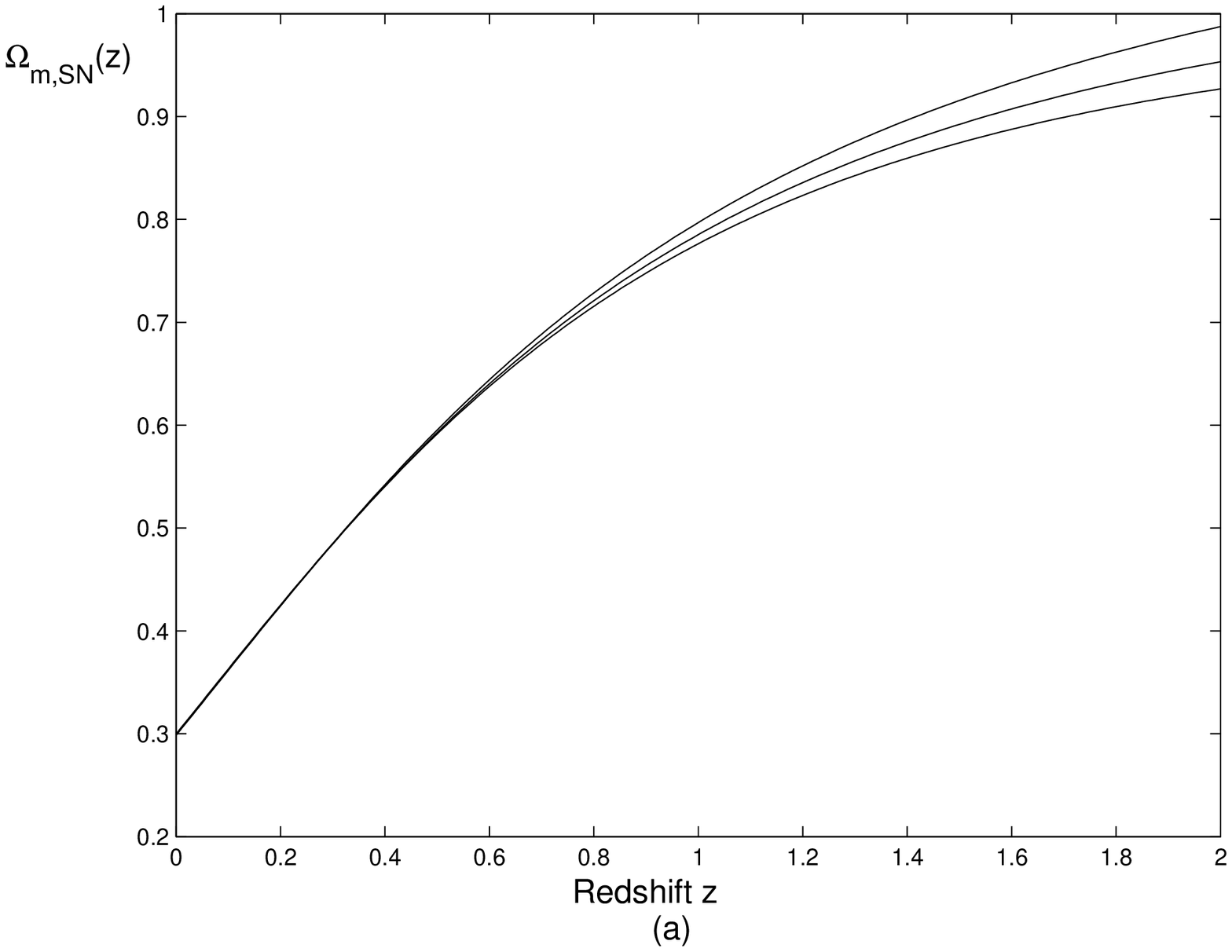}
  \includegraphics[width=0.8\columnwidth]{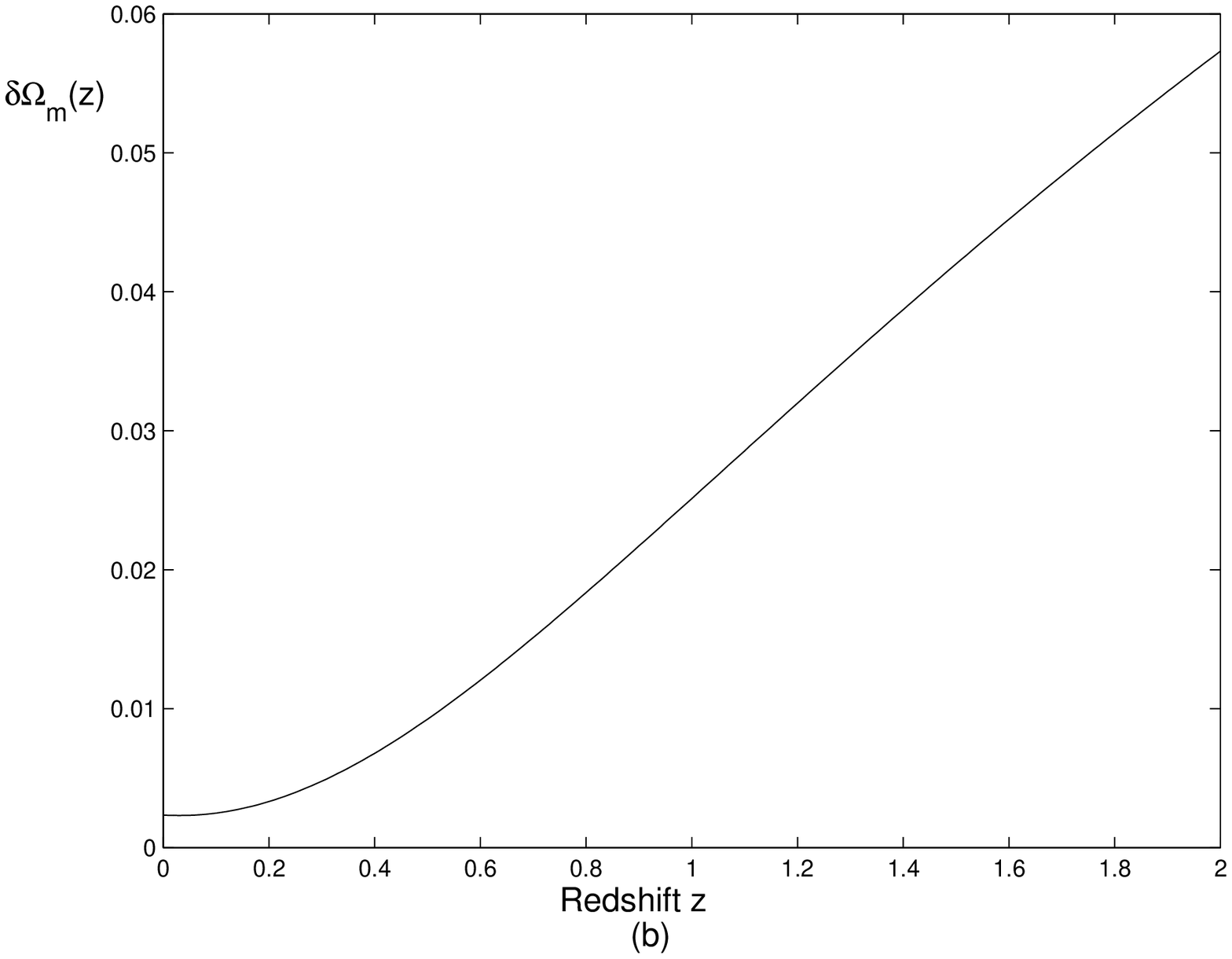}
  \caption{(a). The evolution of the predicted matter energy parameter for $\epsilon=0.01,0.05,0.1$ from bottom to top.
  (b). The fractional change to the standard $\Lambda$CDM with $\epsilon=0.078$.}\label{1}
\end{figure}

It is also necessary to analyze carefully various methods of
constraining matter energy density parameter in $\Lambda(t)$CDM
(see, e.g., Ref.\cite{white} for a review). Due to the modified
matter expansion rate, they may actually constrain different
things in $\Lambda(t)$CDM. For example, if we want to constrain
the evolution of $\Omega_m$ using SNe Ia data, we must first
specify a model for the dark energy. Current popular choice is
take the $\Lambda$CDM model, i.e. we will assume the matter
expands in the standard rate and the cosmological constant is a
true constant. So in SNe Ia observation, if we want to compare the
prediction of the $\Omega_m$ in $\Lambda(t)$CDM with the one in
$\Lambda$CDM, the correct quantity is $\Omega_{m,
SN}(z)=\Omega_{m0}(1+z)^3/(H/H_0)^2$. Fig.3(a) shows the evolution
of $\Omega_{m, SN}(z)$ up to redshift 2 for $\epsilon=0.01, 0.05,
0.1$ from bottom to top with $\Omega_{m0}=0.3$. It is different
from Fig.2 in Ref.\cite{renorm} since they used the definition
$\Omega_m(z)=\Omega_{m0}(1+z)^{3-\epsilon}/(H/H_0)^2$.
Correspondingly, the deviation parameter $\delta\Omega_m$ in our
case, defined as the fractional change of $\Omega_m$ with respect
to the standard $\Lambda$CDM will be different from that of
Ref.\cite{renorm}. This is shown in Fig.3(b) corresponding to the
same $\epsilon$ value of Ref.\cite{renorm}: $\epsilon=0.078$. We
can see that at redshift 1.5, the fractional change is less than
$4\%$ and at redshift 1 it is less than $2.5\%$. In
Ref.\cite{renorm}, those two numbers are $20\%$ and $10\%$
respectively, which is much larger than ours. So we conclude that
for $\epsilon=0.078$, it will be very hard to distinguish it from
$\Lambda$CDM by the evolution of $\Omega_m$ using SNe Ia data in
the near future.

On the other hand, the physical matter energy density $\rho_m$
given by Eq.(\ref{5}) is constrained in more direct methods such
as galaxy dynamics or gravitational lensing. However, since vacuum
is constantly decaying to matter, care should be paid to analyze
in those direct methods what is the redshift that the matter
density is constrained. In this work, since our main interest is
constraining the vacuum decay rate $\epsilon$, in the following
discussion we will simply take $\Omega_{m0}=0.3$.

Constraint on the dark energy EOS can be considerably tightened if
we combine data from SNe Ia, CMB and LSS observations. For
example, in Ref.\cite{ywang}, Yun Wang \emph{et al.} showed that
when combining data from SNe Ia, CMB and LSS, we have
$-1.24<\bar\omega_X<-0.74$ at $95\%$ C.L, which is much tighter
than the bound we quoted above using only the SNe Ia data.
However, in $\Lambda(t)$CDM, it is problematic to apply
constraints on $\bar\omega_X$ from CMB or LSS data. The reason it
that, from Eq.(\ref{eos}), we can see that the effective EOS will
tend to negative infinity at redshift $z_*$ defined by
\begin{equation}
{3\over
3-\epsilon}(1+z_*)^{3-\epsilon}-(1+z_*)^3+\tilde\Omega_{\Lambda0}/\Omega_{m0}=0\
.\label{}
\end{equation}
For example, for $\epsilon=0.01, 0.05, 0.1$, we have $z_*\sim4.6,
2.6, 2.1$, respectively. This pathological behavior of $\omega_X$
is not a sign that some unphysical thing happens at $z_*$. It is
just due to the implicit assumption underlying the definition
(\ref{3}): the modification term to the standard matter dominated
Friedmann equation $\delta H^2$ must be always non-zero. This
means that the expansion rate of the Universe should be always
larger than the standard matter dominated case even at high
redshift. If this assumption fails, i.e. $\delta H^2\rightarrow 0$
as the redshift approaches a finite value $z_*$, then $\omega_{X}$
will tend to infinity at $z_*$. In $\Lambda(t)$CDM, this
divergence in $\omega_X$ is the direct result of the decreased
matter expansion rate. Due to this pathological behavior, if we
consider CMB and LSS data, i.e. we are considering $\bar\omega_X$
averaged up to the redshift of recombination, it is suspicious
that this makes sense.

However, based on the same reasoning of effective EOS formalism,
we can construct the effective energy density of dark energy and
in this formalism we will not encounter the problems in the
effective EOS formalism. So we can compare the effective energy
density to the recent reconstructed dark energy density using SNe
Ia, CMB and LSS data \cite{ywang}.

\begin{figure}
 \includegraphics[width=0.8\columnwidth]{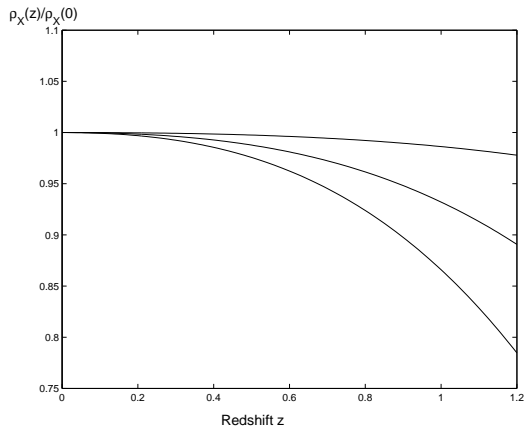}
 \caption{The effective energy density given by Eq.(\ref{den}). The three lines correspond to $\epsilon=
  0.01, 0.05, 0.1$ from top to bottom.}\label{1}
\end{figure}

So let's write the Friedmann equation (\ref{7}) into the form
\begin{equation}
{H^2\over
H_0^2}=\Omega_{m0}(1+z)^3+(1-\Omega_{m0}){\rho_X(z)\over\rho_X(0)}\
,\label{8}
\end{equation}
where $\rho_X$ is the effective energy density of the dark energy
which is given by
\begin{equation}
\rho_X(z)={3\rho_{m0}\over3-\epsilon}(1+z)^{3-\epsilon}-\rho_{m0}(1+z)^3+\tilde\rho_{\Lambda
0}\ .\label{den}
\end{equation}

Fig.4 shows the evolution of $\rho_X$ up to redshift 1.2. This can
be compared to Fig.3(a) of Ref.\cite{ywang2}, which is the most
up-to-date constraint on dark energy density using data from SNe
Ia, CMB and LSS observations. We can see that all three curves in
Fig.4 is consistent with current observational constraint at
$2\sigma$ level.

To confront $\Lambda(t)$CDM with the matter power spectrum of LSS,
it is necessary to consider the evolution of linear perturbations.
It is possible that this will give a tighter constraint on the
vacuum decay rate.

Following from the property that vacuum do not fluctuate, we can
see that the density perturbation $\delta\rho_m=\rho_m-\bar\rho_m$
will satisfy the standard continuity equation,
\begin{equation}
\dot{\delta\rho_m}+3\rho_m\delta H+3H\delta\rho_m=0\ .\label{}
\end{equation}

\begin{figure}
  \includegraphics[width=0.8\columnwidth]{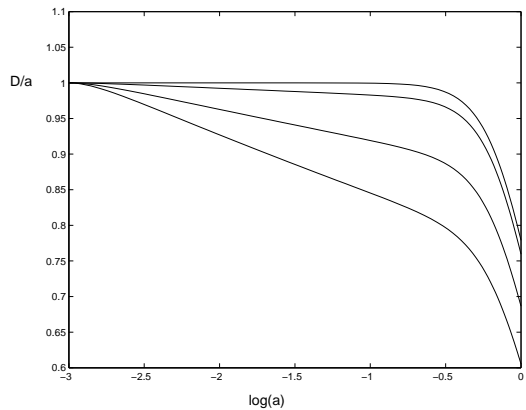}
  \caption{The evolution of linear perturbations for $\epsilon=0.1, 0.05, 0.01,0$ from bottom to top.}\label{1}
\end{figure}

Then following the standard treatment (see, e.g.,
Ref.\cite{liddle}), we can find that the density contrast
$\delta\equiv\delta\rho_m/\bar\rho_m$ satisfies,
\begin{equation}
\ddot\delta+2H\delta-4\pi G\bar\rho_m\delta=0\label{}
\end{equation}
This is just the same form as that in $\Lambda$CDM. In matter
dominated era of $\Lambda$CDM, it is well-known that the density
contrast evolves as $\delta\propto a$ \cite{liddle}. In
$\Lambda(t)$CDM, during matter dominated era, it can be found that
\begin{equation}
\delta\propto a^{(\sqrt{\epsilon^2-6\epsilon+25}-\epsilon-1)/4}\
,\label{per}
\end{equation}
which will be slower than the standard case when $\epsilon>0$.
Thus it will be harder for large scale structure to form in
$\Lambda(t)$CDM.

Fig.5 shows the evolution of the linear growth factor $D\equiv
\delta(a)/\delta(a_i)$ for $\epsilon=0.1, 0.05, 0.01$ from bottom
to top with $\Omega_{m0}=0.3$. The topmost curve corresponds to
the evolution in standard $\Lambda$CDM model. Several features are
quite obvious from this figure. First, in matter dominated era,
the perturbation will evolve slower when $\epsilon$ is larger.
Second, in vacuum dominated era, they evolve in almost the same
rate since the Universe expands exponentially in all those cases.
So under the same initial condition, the growth factor will be
smaller when $\epsilon$ is larger. To say it in other words, for
the same $\sigma_8$, $\Lambda(t)$CDM with larger $\epsilon$ demand
larger density fluctuations at early times. This effect should
lead to a difference between CMB normalization and low-redshift
normalization. For example, we can see that for the case of
$\epsilon=0.05$, the fractional change with the $\Lambda$CDM case
is about $12\%$. Such a non-negligible fractional change can be
distinguished from precise LSS data. Compared to the upper bound
from SNe Ia observation, $\epsilon<0.3$ \cite{renorm}, we can see
that the evolution of linear growth factor can give tighter
constraint on the vacuum decay rate. Finally, all the four
evolution curves begin to slow down at roughly the same redshift.
This is due to the observation in in Sec.\ref{evolution} that, for
$\epsilon<0.1$, the Universe begins to accelerate at roughly the
same redshift.

\section{Conclusions and discussions}

In sum, we discussed the cosmological evolution of decaying vacuum
cosmology based on a simple assumption about the modified matter
expansion rate. We showed that all the existing models of decaying
vacuum cosmology can be elegantly unified in this framework. So
even if we do not understand the physics of vacuum decay, we can
still faithfully talk about whether vacuum is decaying by
constraining the vacuum decay rate using the system
(\ref{5}-\ref{dece}). Based on this approach, we also proposed a
new vacuum decay law. We emphasized that due to the modified
matter expansion rate, it should be careful to make predictions of
cosmological parameters in $\Lambda(t)$CDM. Specifically, we
discussed the effective EOS and density parameter. From the
effective EOS, we indicated that $\Lambda(t)$CDM is a possible
candidate for phantom energy. However, there is a possible trouble
in the effective EOS formalism so we constructed the effective
energy density, from which we showed that $\epsilon<0.1$ is
consistent with current data from SNe Ia, CMB and LSS
observations. Finally, we discussed the evolution of linear
perturbations in $\Lambda(t)$CDM and showed that it can give
tighter constraint than just the cosmological evolution.

\section*{ACKNOWLEDGEMENTS}

P.W. would like to thank Sergei D. Odintsov for helpful comment on
the manuscript and Yun Wang for clarifying correspondence on
current observational constraint on dark energy. X.H.M. would like
to thank the Physics Department of UoA for hospitality. This work
is supported partly by an ICSC-World Laboratory Scholarship, a
China NSF and Doctoral Foundation of National Education Ministry.

\end{document}